\documentclass[12pt]{article}

\begin{document}
\begin{center}
{\bf On Possible Light-Torsion Mixing in Background Magnetic Field}

\vspace{5mm}
 S. I. Kruglov\\

\vspace{5mm}
\textit{University of Toronto at Scarborough,\\ Physical and Environmental Sciences Department, \\
1265 Military Trail, Toronto, Ontario, Canada M1C 1A4}
\end{center}

\begin{abstract}
The interaction of the light with propagating axial torsion fields
in the presence of an external magnetic field has been
investigated. Axial torsion fields appearing in higher derivative
quantum gravity possess two states, with spin one and zero, with
different masses. The torsion field with spin-0 state is a ghost
that can be removed if its mass is infinite. We investigate the
possibility when the light mixes with the torsion fields resulting
in the effect of vacuum birefringence and dichroism. The
expressions for ellipticity and the rotation of light polarization
axis depending on the coupling constant and the external magnetic
field have been obtained.
\end{abstract}

\section{Introduction}

Torsion fields can exist as possible fields in the gravity theory.
In the Einstein-Cartan theory, the simplest generalization of
General Relativity (GR), torsion fields do not propagate
\cite{Hehl}, i.e. they are non-dynamic fields. The first version
of quantum gravity with higher derivatives \cite{Stelle} did not
include torsion. Kinetic terms for torsion, in the framework of
quantum gravity, were considered in many papers (see
\cite{Buchbinder}). GR is a classical theory, but the source of
torsion fields is the spin of matter fields - pure quantum
characteristic. Therefore, torsion fields should be investigated
in the quantum version of gravity.

GR possesses difficulties such as a problem of cosmological
singularity. At the initial time, the solution of GR equations has
the singular state with divergent energy density and singular
metrics. Torsion fields, if they exist, can play an important role
in cosmology, especially in the physics of the early universe. Some
schemes with torsion lead to the singularity-free
cosmological models \cite{Minkevich}. It should be noted that this
property is model dependent.

The Poincar\'{e} gauge theory of gravity also possesses the
Rimann-Cartan structure. At extremely high energy density, the
dynamics depends on the space-time torsion. The absence of a
singularity is connected with gravitational repulsion effect which
is due to the presence of torsion. Possibly, the space-time torsion
could solve the dark energy/matter problem \cite{Minkevich}. Indeed,
dark energy is introduced in cosmology to explain the acceleration
of the cosmological expansion at the present time. The presence of dark
energy leads to a negative pressure. But even without dark energy,
the torsion fields result in the gravitational repulsion effect.

It has to be mentioned that a torsion field also arises in
superstring theory \cite{Polchinski} and in higher dimensional theories
\cite{German}. All this shows the importance to investigate
quantum torsion fields and their interactions.

The metric and torsion are independent fields. Contrarily, in the teleparallel gravity, the curvature and
torsion are alternative ways of describing the gravitation fields.
The classification of torsion fields is given in \cite{Hehl1}.

In this work, we take into account the antisymmetric part of the torsion tensor
$S_{\mu\nu\alpha}$ because only the axial vector
$S^\mu=(1/3!)\epsilon^{\mu\nu\alpha\beta}S_{\nu\alpha\beta}$
interacts with matter minimally \cite{Shapiro}. The importance of
the completely antisymmetric Cartan tensor in Einstein-Sciama-Kibble
theory and Kaluza-Klein multidimensional theories was discussed in
\cite{Fabbri}.

The interaction of torsion fields with fermions was studied in \cite{Shapiro1},
\cite{Kostelecky1}. It was demonstrated in \cite{Shapiro1}, that Large Hadronic Collider
(LHC) can test the torsion-fermion interaction parameters. In \cite{Shapiro1}, constraints
on some constants of the torsion-fermion interaction and the torsion mass were obtained.
It was noticed in \cite{Lammerzahl} (see also discussions in \cite{Shapiro})
that the background torsion breaks the local Lorentz
invariance. Authors \cite{Kostelecky1} found constrains on many torsion components from
the Lorentz violation bounds.

The paper is organized as follows. In Sec. 2, we considered the interaction
of the light with propagating axial torsion fields
in the presence of an external magnetic field. Dispersion relations and solutions
of field equations are obtained. Two cases are examined: the mass of the ghost
if finite, and infinite. The effect of vacuum birefringence and dichroism
is investigated in Sec. 3. In Sec. 4, we discuss the possible observation of the phenomena
of vacuum birefringence and dichroism and make a conclusion.

We considered here the flat Minkowski space-time with torsion and use
the Euclidian metric. The four-vector of the torsion field is
$S_\mu=(S_m,S_4)$; $m=1,2,3$; $S_4=iS_0$. The rationalized Heaviside-Lorentz
units are explored here, and $\hbar=c=1$.

\section{Field Equations and Solutions}

The torsion Lagrangian which appears
in the higher derivative quantum gravity is given
by \cite{Buchbinder}, \cite{Belyaev}:
\begin{equation}
\mathcal{L}_T=-aS_{\mu\nu}^2-b(\partial_\mu
S_\mu)^2 -c S_\mu ^2,
  \label{1}
\end{equation}
where $S_{\mu\nu}=\partial_\mu S_\nu-\partial_\nu S_\mu$. It was
shown in \cite{Kruglov} (see also \cite{Kruglov2}) that the
torsion field possesses two spins, one and zero, with different
masses, so that
\[
a=\frac{1}{4},~~~b=\frac{m^2}{2m_0^2},~~~c=\frac{1}{2}m^2,
\]
and $m$, $m_0$ are masses of spin-1 and spin-0 states, correspondingly.
The spin-0 is a ghost resulting in the negative contribution to the energy,
and one should introduce indefinite metrics in quantum field theory \cite{Kruglov2}.
But at $m_0\rightarrow \infty$ the second term in Eq.(1)
vanishes and one comes to the Proca theory describing a pure spin-1 field.
It has to be mentioned that in the quantum gravity the second term in Eq.(1) appears
due to quantum corrections and we can not remove it by ``hands".
The main difficulty of the quantum field theory with indefinite metrics is its
nonunitarity and this is an attribute of the renormalized quantum gravity \cite{Stelle}, \cite{Buchbinder}. However at a large mass of the ghost compared with the mass of a spin-1
state, one has no difficulties. The relation between the mass of the ghost and unitarity
has been discussed in \cite{Berredo}.
Now we introduce an interaction of vector field with electromagnetic fields (light) in the same manner as an axion interaction.
Thus, we consider the interaction Lagrangian
\begin{equation}
\mathcal{L}_{int}=\frac{1}{4}g\left(\partial_\alpha
S_\alpha\right)\mathcal{F}_{\mu\nu}\widetilde{\mathcal{F}}_{\mu\nu},
  \label{2}
\end{equation}
where $g$ is a coupling constant with the dimension (mass)$^{-2}$, $\mathcal{F}_{\mu\nu}$ is the strength of
electromagnetic fields and $\widetilde{\mathcal{F}}_{\mu\nu}$ is its dual tensor. The Lagrangian (2) is Lorentz and CP invariant. Contrarily, authors \cite{Colladay} investigated a general
Lorentz-violating two-photon interactions.
It should be mentioned that the Lagrangian (2) within four-divergence can be represented as
\[
\mathcal{L}'_{int}=-\frac{1}{4}g
S_\alpha\partial_\alpha\left(\mathcal{F}_{\mu\nu}\widetilde{\mathcal{F}}_{\mu\nu}\right),
\]
such that equations of motion remain the same. It is obvious from this expression that the interaction
introduced can be considered also for the case of pure spin-1 field when the mass of spin-0 approaches to
infinity, $m_0\rightarrow \infty$. We investigate the general case of the presence of spin-1 as well as spin-0 states of the torsion field. One can consider this case as the vector field interaction in the general gauge. At the end of calculations, we may put $m_0\rightarrow\infty$ to have the consistent theory. At the same time, there is a possibility to consider the ghost to be presented in the theory.

Equations of motion, obtained from the Lagrangian $\mathcal{L}=\mathcal{L}_T +\mathcal{L}_{int}
+\mathcal{L}_{el}$, where $\mathcal{L}_{el}=-(1/4)\mathcal{F}_{\mu\nu}^2$ is the Lagrangian of free electromagnetic fields, are given by
\[
\partial_\mu \mathcal{F}_{\mu\nu}=g\partial_\mu\left[\left(\partial_\rho S_\rho\right)\widetilde{\mathcal{F}}_{\mu\nu}\right],
\]
\vspace{-8mm}
\begin{equation}
\label{3}
\end{equation}
\vspace{-8mm}
\[
\partial_\mu^2 S_\nu-\delta\partial_\nu\partial_\mu S_\mu
-m^2S_\nu=\frac{1}{4}g\partial_\nu\left(\mathcal{F}_{\alpha\beta}\widetilde{\mathcal{F}}_{\alpha\beta} \right),
\]
where
\[
\delta=1-\frac{m^2}{m_0^2}.
\]
We examined the case of a propagation of a light field $f_{\mu\nu}$ in the external background constant and uniform magnetic field $F_{\mu\nu}$ ($F_{\mu 4}=0$, $\overline{B}_i=(1/2)\varepsilon_{ijk}F_{jk}$).
Thus, we have
\begin{equation}
\mathcal{F}_{\mu\nu}=f_{\mu\nu}+F_{\mu\nu},
\label{4}
\end{equation}
where $f_{\mu\nu}=\partial_\mu A_\nu-\partial_\nu A_\mu$, and $A_\mu$ being the four-vector potential of
the light field. Now, we specify that the light beam propagates in the z-direction and the
background magnetic field points in the x-direction. With the help of a gage $A_3=A_4=0$, leaving only linear terms in $\textbf{A}$ and $S_\mu$, we find from Eq.(3):
\begin{equation}
\partial_\mu^2 A_1=\beta\partial_t\partial_\rho S_\rho,
\label{5}
\end{equation}
\begin{equation}
\partial_\mu^2 A_2=0,
\label{6}
\end{equation}
\begin{equation}
\partial_\mu^2 S_\nu-\delta\partial_\nu\partial_\mu S_\mu
-m^2S_\nu=\beta\partial_\nu\partial_t A_1,
\label{7}
\end{equation}
where $\beta=g\overline{B}$.
It follows from Eq.(6) that only the component of a light $A_1$ parallel to background magnetic field $\overline{\textbf{B}}$ interacts with the torsion field. Eq.(6) has a solution $A_2=\exp[-i\omega (t-z)]$, so that the index of refraction of a vector-potential component perpendicular to the magnetic field is $n_\bot=1$. Implying the dependence of propagating fields on the $t$ and $z$, we see from Eq. (7) that
$S_1$ and $S_2$ components of the torsion field obey equations for free fields. One can obtain solutions to Eq.(5),(7) of the form
\[
A_1(z,t)=A\exp[-i(\omega t-kz)],
\]
\begin{equation}
S_3(z,t)=S_3\exp[-i(\omega t-kz)],
\label{8}
\end{equation}
\[
S_0(z,t)=S_0\exp[-i(\omega t-kz)].
\]
Replacing Eq.(8) into Eq.(5),(7), we arrive at the system of linear equations
\[
\left(\omega^2-k^2\right)A=\beta\omega\left(kS_3-\omega S_0\right),
\]
\begin{equation}
\left(\omega^2-k^2-m^2\right)S_3-\delta k\left(\omega S_0-kS_3 \right)=\beta\omega kA,
\label{9}
\end{equation}
\[
\left(\omega^2-k^2-m^2\right)S_0-\delta \omega\left(\omega S_0-kS_3 \right)=\beta\omega^2 A.
\]
Now we discuss two cases: the mass of the ghost if finite, and infinite.

\subsection{The ghost mass, $m_0$, is finite}

The homogeneous linear equations (9) possess a solution when the corresponding determinant equals zero. Thus, we obtain the dispersion relation
\begin{equation}
\left(k^2-\omega^2\right)\left(k^2-\omega^2+m^2\right)\left[m^2
\left( k^2-\omega^2+m_0^2\right)-\beta^2\omega^2m_0^2 \right]=0.
\label{10}
\end{equation}
When the background magnetic field vanishes, $\beta=0$, one arrives at the expected dispersion relations for free fields. From Eq.(10), we find three solutions
\begin{equation}
k_1^2=\omega^2,~~ k_2^2=\omega^2-m^2,~~
k_3^2=\omega^2-m_0^2\left(1-\frac{\beta^2\omega^2}{m^2} \right).
\label{11}
\end{equation}
The general solution to Eq.(9) in the magnetic field is given by
\[
A_1(z,t)=e^{-i\omega t}\left(A^{(1)}e^{ik_1z}+A^{(2)}e^{ik_2z}+A^{(3)}e^{ik_3z}\right),
\]
\vspace{-8mm}
\begin{equation}
\label{12}
\end{equation}
\vspace{-8mm}
\[
S_{0,3}(z,t)=e^{-i\omega t}\left(S^{(1)}_{0,3}e^{ik_1z}+S^{(2)}_{0,3}e^{ik_2z}+S^{(3)}_{0,3}e^{ik_3z}\right).
\]
Now we consider a case when the eigenstates having $k_2$ and $k_3$ are not degenerated, i.e.
$k_2\neq k_3$. Replacing Eq.(12) into equations of motion (5)-(7), one obtains the relations at
$k_2\neq k_3$ (or $\beta^2 \omega^2 \neq m^2\delta$):
\begin{equation}
A^{(1)}=-\frac{m^2}{\omega^2\beta}S^{(1)}_0,~~~~S^{(1)}_0=S^{(1)}_3,
\label{13}
\end{equation}
\begin{equation}
A^{(2)}=0,~~~~k_2S^{(2)}_3=\omega S^{(2)}_0,
\label{14}
\end{equation}
\begin{equation}
A^{(3)}=-\frac{\beta\omega}{k_3}S^{(3)}_3,~~~~\omega S^{(3)}_3=k_3S^{(3)}_0.
\label{15}
\end{equation}
Then Eq.(12) become
\begin{equation}
\left(\begin{array}{c}A_1(z,t)\\S_0(z,t)\\S_3(z,t)
\end{array}\right)=e^{-i\omega t}\biggl[\left(
\begin{array}{c}m^2\\-\omega^2\beta\\-\omega^2\beta \end{array}\right)ae^{ik_1z}+
\left(\begin{array}{c}0\\\omega k_2\\\omega^2 \end{array}\right)be^{ik_2z}+
\left(\begin{array}{c}-\beta \omega^2\\\omega^2\\\omega k_3 \end{array}\right)ce^{ik_3z}\biggr],
\label{16}
\end{equation}
where $a$, $b$, $c$ are normalization constants. It follows from Eq.(11) that at the limit $m_0\rightarrow \infty$, the momentum $k_3$ formally should take the limit $k_3\rightarrow i\infty$. As a result, the last exponential factor in Eq.(16) vanishes, and the photon field component $A_1(z,t)$ (as well as $A_2(z,t)$) does not mix with the torsion field and index of refraction of a vector-potential component parallel to the magnetic field is $n_\|=1$. Therefore, in this case there are no effects of vacuum birefringence and dichroism. Thus, the ghost associated with the scalar part of $S_\mu$ is decoupled. This case is not interesting for us because we look for the phenomenon of vacuum birefringence and dichroism.

Let us consider the case $\beta^2 \omega^2 = m^2\delta$ ($k_2= k_3$). Remember that at the limit $m_0\rightarrow \infty$, we have $\delta\rightarrow 1$. Now, we treat the relation $\beta^2 \omega^2 = m^2\delta$ as the fine tuned case that can happen by varying $\omega$ or the magnetic field entering in $\beta=g\overline{B}$. Thus, this is not a condition for the mass $m$ of a spin-1 state of the torsion field. Equations of motion (9) lead to constrains on the integration constants. Eq.(13) is valid for this case, but new constraint is:
\begin{equation}
A^{(2)}=\frac{\omega\beta}{m^2}\left(k_2S^{(2)}_{3}-\omega S^{(2)}_{0}\right).
\label{17}
\end{equation}
In this degenerated case, there are two independent variables $S^{(2)}_{3}$, $S^{(2)}_{0}$, and
the solution is given by
\begin{equation}
\left(\begin{array}{c}A_1(z,t)\\S_0(z,t)\\S_3(z,t)
\end{array}\right)=e^{-i\omega t}\biggl[\left(
\begin{array}{c}m^2\\-\omega^2\beta\\-\omega^2\beta \end{array}\right)ae^{ik_1z}+
\left(\begin{array}{c}\omega\beta\left(k_2S^{(2)}_{3}-\omega S^{(2)}_{0}\right)\\m^2S^{(2)}_{0} \\m^2S^{(2)}_{3} \end{array}\right)e^{ik_2z}\biggr].
\label{18}
\end{equation}
To remove the ghost, one may consider here the constrain $m_0\rightarrow\infty$, $\delta=1$.
As a result, to have the phenomenon of vacuum birefringence and dichroism, we must examine
the fine tuned case $m=\omega\beta$. We will analyze this case later on.

\subsection{The ghost mass is infinite, $m_0\rightarrow\infty$}

In this case the ghost is absent from the very beginning, and we discuss the limit
$m_0\rightarrow\infty$ in Eq.(1), (3), (9). Then $\delta=1$, and Eq.(9) become
\[
\left(\omega^2-k^2\right)A=\beta\omega\left(kS_3-\omega S_0\right),
\]
\begin{equation}
\left(\omega^2-m^2\right)S_3- k\omega S_0=\beta\omega kA,
\label{19}
\end{equation}
\[
-\left(k^2+m^2\right)S_0+\omega kS_3 =\beta\omega^2 A.
\]
The homogeneous linear equations (19) possess non-trivial solutions when the
determinant equals zero:
\begin{equation}
\det\left(
\begin{array}{ccc}
\omega^2-k^2 & \omega^2\beta & -\omega k\beta\\
\omega k\beta & \omega k & m^2-\omega^2 \\
\omega^2\beta & m^2+k^2 & -\omega k
\end{array}
\right)=0.
\label{20}
\end{equation}
From Eq.(20), we arrive at the dispersion relation
\begin{equation}
\left(k^2-\omega^2\right)\left(k^2-\omega^2+m^2\right)\left(m^2
-\omega^2\beta^2\right)=0.
\label{21}
\end{equation}
Thus, the general solution to Eq.(19) is
\[
A_1(z,t)=e^{-i\omega t}\left(A^{(1)}e^{ik_1z}+A^{(2)}e^{ik_2z}\right),
\]
\vspace{-8mm}
\begin{equation}
\label{22}
\end{equation}
\vspace{-8mm}
\[
S_{0,3}(z,t)=e^{-i\omega t}\left(S^{(1)}_{0,3}e^{ik_1z}+S^{(2)}_{0,3}e^{ik_2z}\right),
\]
where $k_1^2=\omega^2$, $k_2^2=\omega^2-m^2$.
Let us examine different solutions. If $m\neq \omega\beta$ (no fine tuning), one
arrives at the trivial solution:
\begin{equation}
A^{(1)}=-\frac{m^2}{\omega^2\beta}S^{(1)}_0,~~~~S^{(1)}_0=S^{(1)}_3,
\label{23}
\end{equation}
\begin{equation}
A^{(2)}=0,~~~~k_2S^{(2)}_3=\omega S^{(2)}_0,
\label{24}
\end{equation}
and Eq.(22) become
\begin{equation}
\left(\begin{array}{c}A_1(z,t)\\S_0(z,t)\\S_3(z,t)
\end{array}\right)=e^{-i\omega t}\biggl[\left(
\begin{array}{c}m^2\\-\omega^2\beta\\-\omega^2\beta \end{array}\right)ae^{ik_1z}+
\left(\begin{array}{c}0\\\omega k_2\\\omega^2 \end{array}\right)be^{ik_2z}\biggr].
\label{25}
\end{equation}
In this case the photon field component $A_1(z,t)$ and the torsion field are not mixed and no effects of vacuum birefringence and dichroism. This corresponds to the limit $m_0\rightarrow\infty$, $k_3\rightarrow i\infty$ in Eq.(16).

For fine tuning case, $m = \omega\beta$, we obtain the solution
\[
A^{(1)}=-\beta S^{(1)}_0=-\beta S^{(1)}_3,
\]
\vspace{-8mm}
\begin{equation}
\label{26}
\end{equation}
\vspace{-8mm}
\[
A^{(2)}=\frac{1}{\omega\beta}\left(k_2S^{(2)}_{3}-\omega S^{(2)}_{0}\right),
\]
which can be represented as follows:
\begin{equation}
\left(\begin{array}{c}A_1(z,t)\\S_0(z,t)\\S_3(z,t)
\end{array}\right)=e^{-i\omega t}\biggl[\left(
\begin{array}{c}\beta\\-1\\-1 \end{array}\right)Ce^{ik_1z}+
\left(\begin{array}{c}k_2S^{(2)}_{3}-\omega S^{(2)}_{0}\\\omega\beta S^{(2)}_{0} \\\omega\beta S^{(2)}_{3} \end{array}\right)e^{ik_2z}\biggr],
\label{27}
\end{equation}
and $C$ is the normalization constant. Eq.(27) is consistent with Eq.(18), and therefore, the ghost can
be removed from the theory by the limit $m_0\rightarrow \infty$. This consideration shows that we can avoid the presence of the ghost by putting from the very beginning $m_0\rightarrow\infty$ in Eq.(1),(3),(9). As a result, at $m=\omega\beta$ there is the effect of vacuum birefringence and dichroism. The same conclusion follows from the degenerated case (with $k_2=k_3$) of subsection 2.1 when the mass of the ghost approaches to infinity at the end of calculations.

\section{Vacuum Birefringence and Dichroism}

Thus, to have the effects of vacuum birefringence and dichroism, we think over
the fine tuned case $m=\omega\beta$,  and $m_0\rightarrow \infty$, $\delta= 1$.
Imposing the initial amplitudes at $t=0$, $z=0$:
\begin{equation}
A_1(0,0)=1,~~~~S_{0,3}(0,0)=0,
\label{28}
\end{equation}
and using the values $m=\omega\beta$, $k_2=\omega\sqrt{1-\beta^2}\simeq \omega\left(1-\beta^2/2 \right)$ (at $\beta\ll 1$), we arrive from Eq.(27) at the unique solution
\begin{equation}
A_1(z,t)=\exp\left(i\omega(z-t)\right)\left[2-\exp\left(-i\frac{\omega \beta^2}{2}z\right)\right],
\label{29}
\end{equation}
\begin{equation}
S_{0,3}(z,t)=\frac{2}{\beta}\exp\left(i\omega(z-t)\right)\left[\exp\left(-i\frac{\omega \beta^2}{2}z\right)-1\right].
\label{30}
\end{equation}
We note that Eg.(29),(30) are approximate equations which were obtained from exact solutions (27) by expanding the $k_2$ in the small parameter $\beta$. One can verify that solutions (29),(30) satisfy Eq.(5),(7) (and Eq.(9)) at the limit $m_0\rightarrow\infty$ ($\delta=1$) with the accuracy of $\textit{O}(\beta^4)$.

Let us consider the linearly polarized photon beam at the angle $\theta$ between the initial polarization vector and the magnetic field at $t=0$, $z=0$:
\begin{equation}
A_1(0,0)=\cos\theta,~~~~A_2(0,0)=\sin\theta,
\label{31}
\end{equation}
with the magnetic field $\overline{\textbf{B}}=(\overline{B},0,0)$. Then at $z=L$, implying the condition $\omega \beta^2L\ll 1$, from Eq.(29), we obtain the approximate photon field
\[
A_1(L,t)=\cos\theta\left[1+i\frac{\omega\beta^2}{2}L+\frac{1}{2}\left( \frac{\omega\beta^2}{2}L\right)^2\right]\exp\left({-i\omega (L-t)}\right),
\]
\vspace{-8mm}
\begin{equation}
\label{32}
\end{equation}
\vspace{-8mm}
\[
A_2(L,t)=\sin\theta \exp\left({-i\omega (L-t)}\right).
\]
In Eq.(32), we left terms up to the second order in the small parameter $\omega \beta^2L$.
From Eq.(32), one arrives at the index of refraction of the component of the light field parallel to the magnetic field
\begin{equation}
n_\parallel =1+\frac{\beta^2}{2}.
\label{33}
\end{equation}
The same situation occurs in the case of photon-axion mixing. It was pointed in \cite{Raffelt} that for P- and C-invariant photon-axion interaction only the $\parallel$ photon state mixes with the axion.
From Eq.(33) and the value $n\bot =1$, we obtain \cite{Kr} the ellipticity (the ratio of the minor to the major axis) as follows:
\begin{equation}
\Psi=\frac{1}{4}\omega \beta^2L\sin2\theta.
\label{34}
\end{equation}
Thus, we find the effect of vacuum birefringence caused by the light-torsion interaction in the background magnetic field. Eq.(32) allows us to find the angle of the rotation of the polarization axis \cite{Maiani}, i.e. vacuum dichroism:
\begin{equation}
\Delta\theta=-\frac{1}{16}\omega^2\beta^4L^2\sin2\theta.
\label{35}
\end{equation}
If the effect of vacuum birefringence and dichrois is discovered, one can find from Eq.(34),(35) bounds on the coupling constant $g$,
and the mass of the torsion field (from the relation $m=\omega\beta$).

\section{Conclusion}

We have considered the axial torsion field possessing two masses,
$m$, $m_0$, with spin one and zero interacting with the light in the external magnetic field.
The spin-0 state is a ghost and we imply that its mass $m_0$ approaches to infinity to avoid nonunitarity
of the theory. It should be stressed also that ghosts (k-essence, Phantom ) are widely used in
modern cosmology \cite{Mukhanov}. In the fine tuned case $m=\omega \beta=\omega g\overline{B}$, we predict the effect of vacuum birefringence and dichroism due to mixing between one photon polarization and the new particle. This is a consequence of the specific interaction of the torsion field with the electromagnetic field. We have discussed the general case of the ghost presence in the theory because ghosts appear due to quantum corrections in quantum gravity. But when the mass of the spin-0 state approaches to infinity, the ghost is removed completely. It is seen from Eq.(2) that we consider a coupling of the two photon vertex with the scalar component of the $S_\mu$ field. Experimentalists can look for a new spin-1 particle, regardless the connection to the torsion field, by the observation of the phenomena of vacuum birefringence and dichroism, varying frequency of a laser beam and the strength of the external magnetic field to reach the case $m=\beta\omega$. For fixed experimental values $\omega$ and $\overline{B}$ the fine tuning condition $m=g\omega\overline{B}$ defines a line in $g$, $m$ space. But if the rotation of the polarization axis or ellipticity are discovered, equations (34), (35) fix the $\beta$ and the coupling constant $g$, and automatically the mass of the torsion field $m=\omega\beta$.

 In this paper, we did not discuss the theoretical values for $g$ and $m$ and other possible experiments that could produce bounds on the coupling considered. The main difference of our Lagrangian (2) from the axion case is that the operator has dimension-6, and we expect the energy dependence to be much stronger. In addition, one may examine other possible operators of the same dimension. It is also very important to obtain bounds coming from collider phenomenology and astrophysics. All these we leave for further investigations.

\end{document}